\documentclass[aps,prl,reprint,groupedaddress,preprintnumbers]{revtex4-1}

\usepackage{mathtools}
\usepackage{amssymb}
\usepackage{eqnarray,amsmath}
\usepackage{txfonts}
\usepackage{bbold}
\usepackage[mathscr]{euscript}
\usepackage{bbm}
\usepackage{verbatim}
\usepackage{stmaryrd}
\usepackage{pgf}
\usepackage{url} 
\usepackage{parskip}
\usepackage[bottom]{footmisc}

\newcommand{\eM}     {$\varepsilonup$-machine}
\newcommand{\eMs}    {$\varepsilonup$-machines}
\newcommand{\EM}     {$\varepsilonup$-Machine}

\newcommand{\CausalState}	{ {\cal S} }

\newcommand{\Cmu}		{ {C_\mu}}
\newcommand{\hmu}		{ {h_\mu}}

\newcommand{\Range} {{r}}

\newcommand{\eMSR}    {{{$\varepsilonup$}MSR}}

\newcommand{\Hagg} {H\"{a}gg}

\newcommand{\ie} {{\it {i.e.}}}
\newcommand{\etal} {{\it {et.~al.}}}

\newcommand{\CHC} {ChC}
\newcommand{\CLC} {ClC}

\begin{document}

\title{Chaotic Crystallography:\\
How the physics of information reveals structural order in materials}

\author{D.~P.~Varn}
\email[]{dpv@complexmatter.org}
\homepage[]{http://wissenplatz.org}
\affiliation{{Complexity Sciences Center\\
	Physics Department\\
	University of California, One Shields Avenue,
	{Davis, California} 95616, {USA}}}

\author{J.~P.~Crutchfield}
\email[]{chaos@ucdavis.edu}
\homepage[]{http://csc.ucdavis.edu/$\sim$chaos/}
\affiliation{{Complexity Sciences Center\\
	Physics Department\\
	University of California, One Shields Avenue,
	{Davis, California} 95616, {USA}}}

\date{\today}

\begin{abstract}
We review recent progress in applying information- and
computation-theoretic measures to describe material structure that transcends
previous methods based on exact geometric symmetries. We discuss the necessary
theoretical background for this new toolset and show how the new techniques
detect and describe novel material properties.  We discuss how the approach
relates to well known crystallographic practice and examine how it provides
novel interpretations of familiar structures. Throughout, we concentrate on
disordered materials that, while important, have received less attention both 
theoretically and experimentally than those with either periodic or aperiodic order. 
\end{abstract}

\preprint{Santa Fe Institute Working Paper 14-09-036}
\preprint{arxiv.org:1409.5930 [cond-mat.stat-mech]}

\maketitle

{\bf \large{{Introduction}}}

It is difficult to exaggerate the importance and influence of crystallography
over the past century. Twenty-nine Nobel prizes have been awarded for
discoveries either in or related to crystallography, with at least one prize
per decade~\cite{Nobel14a}. Crystallography strongly influences and is
influenced by other fields, such as chemistry, biology, biochemistry, physics,
materials science, mathematics, and geology, making it perhaps the
quintessential interdisciplinary science~\footnote{Mackay~\cite{Mack86a,Cart12a} shows
a `concept-association network' of research areas as they relate to
classical crystallography. There are considerable connections with
other seemingly disparate fields.}. So ingrained in other disciplines, it is
now often thought of as a service science, in the sense that the techniques and
theory developed in crystallography have become standard tools for researchers
in these other fields. Often among the first questions in a research problem
is `What is the crystal structure of this material?'---or, more
colloquially---{\emph{`Where are the atoms?'}} 

Unquestionably crystallography is a mature field.  The {\emph {International
Tables for Crystallography}} consists of eight volumes (A-G, A1) and if printed
out would, collectively, require nearly 6000 pages~\cite{URL_PagesInITCR}.
Together they coalesce and codify the combined knowledge of the worldwide
crystallographic community.  Additionally, there are at least a dozen major
crystallographic databases, some cataloging hundreds of thousands of different
solved crystal structures~\footnote{A few examples are the Crystallography Open
Database ({{\protect\url{http://www.crystallography.net/}}}), the Cambridge
Structural Database
({{\protect\url{http://www.ccdc.cam.ac.uk/pages/Home.aspx}}}), Pearson's
Crystal Data ({{\protect\url{http://www.crystalimpact.com/pcd/Default.htm}}}),
and the Worldwide Protein Data Bank ({{\protect\url{http://www.wwpdb.org/}}}).
A list of databases is maintained by the International Union of
Crystallographers at {{\protect\url{http://www.iucr.org/resources/data}}}.}
with tens of thousands being added yearly.

As successful as this research program has been, there has been an inordinate interest in those material
structures that possess {\emph {periodic order}} and
thus have discrete reflections in their diffraction patterns, called Bragg peaks~\footnote{
              In fact, until 1992 the defining feature of a
              crystal was the presence of periodic order, when this definition was changed 
              due to the discovery of quasicrystals~\cite{Shec84a}.
              Now, any specimen that has {``}an essentially sharp diffraction 
              pattern{"}~\cite{URL_DefintionOfCrystal92,URL_DefintionOfCrystal} is
              officially classified as a crystal. Nonetheless, when we use the
              term `crystal', we mean materials with periodic order}.
Even in the early days of X-ray crystallography, though, some materials were
known to have considerable diffuse scattering between the Bragg peaks~\cite{Welb14a} or even
to lack Bragg peaks altogether~\cite{Guin63a}. While an observed broadband
spectrum is sometimes a result of thermal agitations or limited experimental resolution, 
it can be and often is a signal of {\emph{disorder}} within the
material. And this disorder can be mild, preserving the integrity of the Bragg
reflections, or it can be severe, where no identifiable long-range order is
present. These cases have not, however, received nearly the same attention as
those with ``an essentially sharp diffraction
pattern"~\cite{URL_DefintionOfCrystal92,URL_DefintionOfCrystal} nor has the
progress been nearly as impressive. Indeed, in some sense disordered structures
have been defined to be outside the 
field of crystallography~\footnote{Mackay in particular has argued that the range of
          crystallography should extend outside its traditional boundaries. {``}Crystallography is only
          incidentally concerned with crystals \dots\ crystallography is rapidly becoming the science of
          structure at a particular level of organization, being concerned with structures bigger 
          than those represented by simple atoms but smaller than those of,
          for example, the bacteriophage. It deals with form and function at those levels, particularly 
          with the way in which large-scale form is the expression of local force.{"}~\cite{Mack75a} 
          }.
Nonetheless crystallographers, defined broadly here as that community of
researchers tasked with understanding and characterizing the atomic arrangement
and composition of materials, have shown a persistent interest in
them~\cite{Guin63a,Seba94a,Welb04a,Egam13a}.

Researchers are increasingly discovering that disorder has profound
effects on material properties and, perhaps surprisingly, disorder can improve
their technological usefulness. For example, it was recently shown that
significantly disordered graphene nanosheets are excellent candidates for use
in high-capacity Li ion batteries due to their unusually high reversible
capacities~\cite{Pan09a}.
Theoretical investigations suggest
that the band gap in ZnSnP$_2$, a promising candidate for high-efficiency
solar cells, changes considerably (0.75 eV -- 1.70 eV) 
as the material transitions from an 
ordered chalcopyrite structure to a disordered sphalerite structure~\cite{Scan12a}.

The growing importance of disorder in materials, then, contrasts sharply with the
lack of tools available to characterize disordered materials.  And, just as
researchers developed new conceptual models and theoretical techniques
to understand the novel organizational structure in quasicrystals~\cite{Jans14a},
new approaches are needed to characterize disordered materials. Here, we detail
a recent initiative that exploits information- and computation-theoretic ideas
to classify the structure of materials in a new way, one that can seamlessly
bridge the gap between perfectly ordered materials, those materials with
some disorder, and finally those that have no discernible underlying crystal structure.\\

{\bf \large{{Classical crystallography}}}

Historically, crystals have been viewed as an unbounded repetition of atoms that
fills 3D space~\footnote{Discussions of these well known concepts from crystallography
                                       are available in any standard text on condensed matter
                                       physics~\cite{Kitt05a,Ashc76a}. For 
                                       a definitive exposition, see the {\emph {International Tables for Crystallography, Vols. A and A1}}.
                                       For classical crystallography, we
									   exclude the case of quasicrystals and,
                                       thus, define a crystal as a periodic arrangement of atoms 
                                       (the pre-1992 definition) rather than by its diffraction pattern (the post-1992 definition).}.
Traditionally one divides this repetition into two parts: the {\emph {basis}}
and the {\emph {lattice}}. The basis is a {\emph {fundamental structural unit}}
composed of one or more atoms. Although the basis can be simple in the extreme,
as for example in Cu, Fe, and alkali metals where there is one atom in the
basis, it can be also much more complicated, as for example in some inorganic
crystals and proteins, where in the latter the basis can be composed of tens of
thousands of atoms. Conversely, the lattice is a mathematical abstraction. It
is defined as a regular periodic collection of points, such that if one
translates from one lattice point to another, the entire arrangement of lattice
points appears to be identical. There are only a finite number of ways that
points can be so distributed in space. In fact, there are fourteen lattice
types in three dimensions (3D), and these are gathered into seven systems:
triclinic, monoclinic, orthorhombic, tetragonal, cubic, trigonal and hexagonal.


To form a {\emph {crystal structure}} then, the basis is attached to each lattice point, 
with each basis having an identical orientation. This is conveniently summarized as~\cite{Kitt05a}:
\begin{align}
   {\mathrm {\bf{crystal \,\,\, structure = basis}}} \times  {\mathrm {\bf{lattice}}}~.
\label{Eq:CrystalStructure}   
\end{align}
Each crystal structure belongs to one of the 230 different {\emph {crystallographic space groups}}, which 
are defined by the symmetries of the crystal, including translations,
rotations, reflections, glides, and screw dislocations. Thus, the regular distribution of matter in space can be classified 
according to physical symmetry operations respected by the crystal structure. So important is
this approach that is has been referred to as {\emph {classical crystallography}} (\CLC)~\cite{Mack86a} 
and may be defined as {\emph {the categorization of material structures based on the 
geometric symmetries respected by the atoms and formally couched in the language of group theory.}}
Succinctly put then, given some material, a primary task of \CLC\ is to
identify the basis and to which of the 230 crystallographic space groups the crystal structure belongs. 
In doing so, \CLC\ provides an answer to the question---{\emph {Where are the atoms?}} \\

{\bf \large{{Towards a new crystallography}}}

The exact symmetries captured by groups fail partially or utterly, however,
depending on a material's degree of disorder. Thus, an alternative is required;
one that naturally adapts to describe randomness and noisy, partial symmetries.

{\bf {Processes defined}}: 
Consider an infinite sequence of random variables, as one might encounter from
time series measurements or as one scans the positions of atoms along one
direction in a material. Formally, we say that there is an ordered sequence of
variables indexed by subscripts and written as $\{ \dots {X}_{-2}, {X}_{-1},
{X}_0,  {X}_{1}, {X}_{2}, \dots \}$. If we make an observation of this
sequence, we observe a specific realization given in lower case: $\{ \dots
x_{-2}, x_{-1}, x_0,  x_{1}, x_{2}, \dots \}$.  We define a {\emph {process}}
as {\emph {the collection of all the possible behaviors that the system may
exhibit, \ie, as the set of all possible realizations of the system.}}
The ensemble of all possible realizations implies a probability distribution
over length-$L$ sequences, at each finite $L$. We will
find that {\emph {identifying the process that describes a material is
analogous to determining the lattice in \CLC.}} We assume that all the
processes considered here are {\emph {stationary}}, 
in the weak sense that their sequence distributions are not functions of
absolute position in space.

\begin{figure*}
\begin{center}
\resizebox{!}{3.9in}{\includegraphics{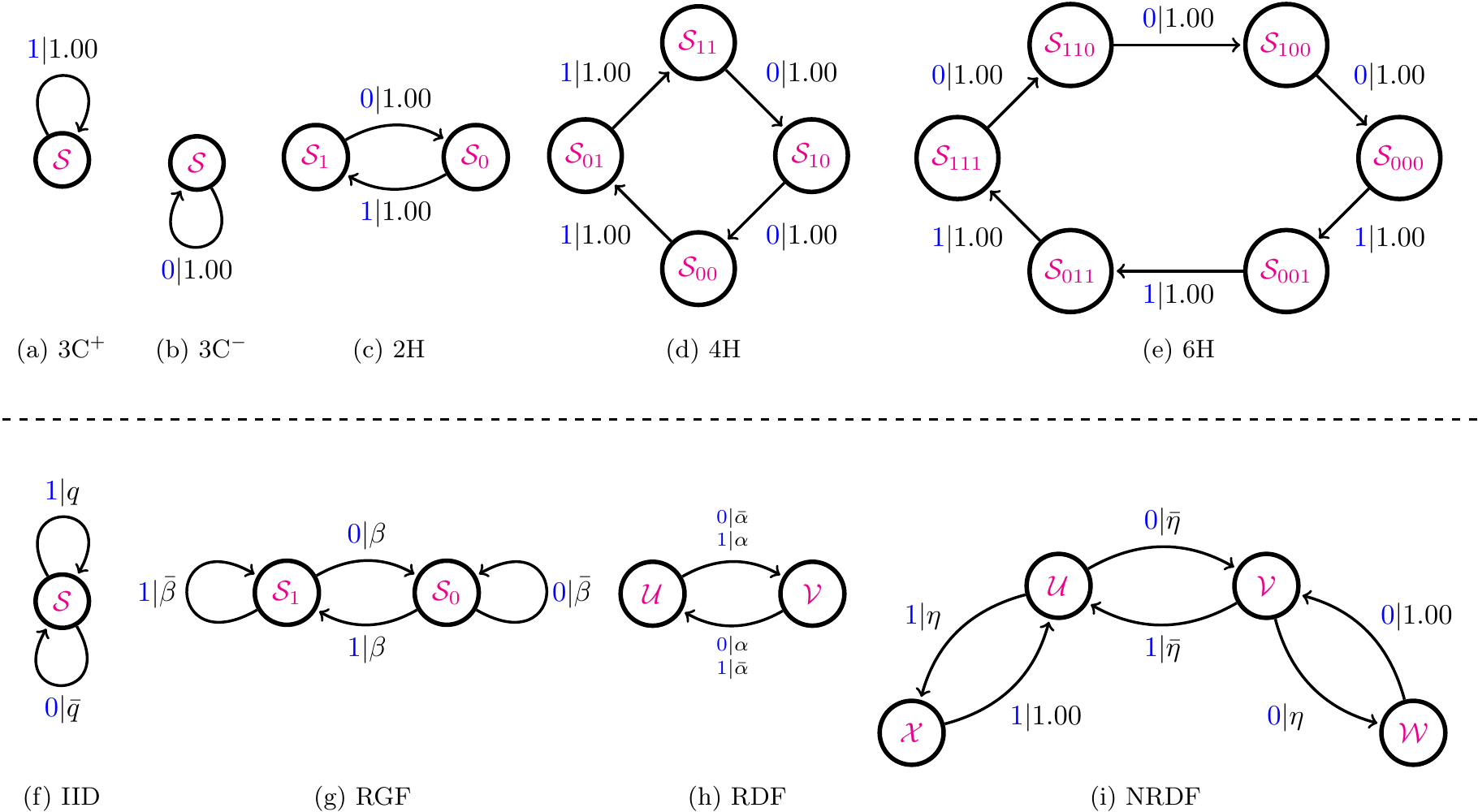}}
\end{center}
\caption{
  Nine \eMs\ that represent ordered (a-e; above the dashed line) and 
  disordered (f-i; below the dashed line) material structures. For each the
  set of output symbols is chosen from $\mathcal{A} = \{0,1\}$. 
  The first seven \eMs, (a-g), are {\emph {finite order Markov processes}}, and the
  CSs are labeled by $\mathcal{S}$ with subscripts giving the minimum number
  of previous symbols necessary to uniquely place the process in that CS. 
  In contrast, the last two \eMs, (h) and (i), may require an indefinitely long history to place them in
  a particular CS. These \eMs\ represent {\emph {strictly sofic}} processes. 
  The CSs are labeled with the 
  symbols $\mathcal{U}, \mathcal{V}, \mathcal {W}, \mathcal{X}$. Arcs connecting
  CSs are labeled $s|p$, where $s$ is the symbol emitted on transition and $p$ is the 
  probability of a transition. A bar over a transition probability is defined as $\bar{p} \equiv 1-p$. 
  (a) 3C$^{+}$ crystal structure.
  (b) 3C$^{-}$ crystal structure. 
  (c) 2H crystal structure. 
  (d) 4H crystal structure. 
  (e) 6H crystal structure. 
  (f) Independent and identically distributed (IID) process~\cite{Cove06a,Riec14b}. For $q=\bar{q} =1/2$, 
       the process is maximally random.  
  (g) Random growth fault (RGF) process. For $\beta$ small, we have a randomly
  twinned 3C structure and,
        for $\beta$ large, there are random growth faults in the 2H structure.
  (h) Random deformation fault (RDF) process. For $\alpha$ small, we have random deformation faulting in 2H.
  (i) Nonrandom deformation fault (NRDF) process. For $\eta$ small, 
       this is nonrandom faulting in 2H; for $\eta$ large, this is a nonrandomly twinned 3C structure.
  }
\label{Fig:CrystalStructures}
\end{figure*}

{\bf {Information theory}}: 
Inherent in the notion of disorder is uncertainty, and the amount of
uncertainty is quantified by {\emph {information
theory}}~\cite{Shan48a,Cove06a}. Imagine a random variable ${X}$ that assumes
discrete outcomes $x \in \mathcal{A}$, where the latter is the set of all
possible outcomes. If before a measurement the result is predicted, then there
is no uncertainty in the outcome and one learns nothing by observing it. If all
possible outcomes are equally likely (maximum ignorance) then, before the
measurement, the result is maximally uncertain and much is learned by
discovering the result.  The genius of Claude Shannon was that this notion can
be quantified and, subject to a few reasonable restrictions, one can define a
unique function (up to an overall scaling factor) that measures the degree of
uncertainty and hence the amount of information learned from a measurement. It
is given by the {\emph {Shannon entropy}} $H[{X}]$ as~\cite{Shan48a,Cove06a}:
\begin{align}
    H[{X}] = -\sum_{x \in \mathcal{A}} \Pr(x) \log_2 \Pr(x)~,
\label{Eq:ShannonEntropy}
\end{align}
where $\Pr(x)$ is the probability of observing a particular realization $x$
when the random variable ${X}$ is measured. If the logarithm is taken to base
2, as is done here, the units of the Shannon entropy are {\emph {bits}}.

Shannon entropy has many multivariate extensions used to capture 
multivariate correlations. In particular, there are the oft-used {\emph {joint entropy}} 
(the Shannon entropy of two or more variables), {\emph {conditional entropy}} 
(the Shannon entropy of a variable conditioned on the outcome of one or more
additional variables), 
and {\emph {mutual information}} (the information shared between two or more variables). 
Other measures have been recently introduced in the literature that identify a 
new range of correlation types~\cite{Jame11a,Jame14a}.

{\bf {Computational mechanics}}: 
There is a well studied theory of correlated, discrete random variables called
{\emph {computational mechanics}}~\cite{Crut89a,Feld98a,Shal01b,Crut12a}.
Within computational mechanics many processes of interest are conveniently
represented as a kind of hidden Markov model~\cite{Rabi89a,Elli95a} known as an
$\varepsilon$-{\emph {machine}}.  In turn, \eMs\ can often be written as directed finite state
automata (FSA)~\cite{Hopc79a}, where the nodes are called {\emph {causal
states}} (CSs) and are connected by directed arcs that represent {\emph
{transitions}} between the CSs. The arcs are labeled $s|p$, where $s$ is the
symbol emitted (observed) upon transition between CSs (which generally are not
directly observable).  The set of CSs, which we denote $\mathbb{S}$, together
with the transition probabilities between them, the set of output symbols
$\mathcal{A}$, and the initial state probability distributions define the \eM.
Critically, instead of being described by group theory, such as one finds in the
crystallographic space groups, the mathematical structure of the \eM\ is that
of a {\emph {semi-group}}. This relaxed mathematical construct allows the \eM\
to capture the approximate symmetries of the process in a natural and
self-consistent manner. This becomes essential in disordered materials, where
strict spatial symmetries may no longer exist.

Importantly, \eMs\ are written using the {\emph {minimal}} number of states,
and all CSs have a unique successor CS upon transition with a particular
symbol, a property called {\emph {unifilarity}}~\cite{Elli09a}.  It can be
shown that the \eM\ for a process is {\emph {unique}}---in the sense that any
other minimal representation is isomorphic to it---and {\emph {optimal}}---in
the sense that no other representation captures more of the
structure~\cite{Shal01b}.  Figure~\ref{Fig:CrystalStructures} shows nine \eMs\
that are important in crystallography.  We call the arrangement of CSs and
their transitions the {\emph {causal architecture}} of the \eM, and the
discovery, study, and interpretation of a process's causal architecture is one
of the main goals of computational mechanics.

{\bf {Measures of intrinsic computation}}:
Glancing at Fig.~\ref{Fig:CrystalStructures}, one notices some obvious
differences between the \eMs: (i) some have more CSs than others
and (ii) some have multiple outgoing transitions for some of their CSs. The first property relates to an intuitive notion
of structure, which can be quantified in terms of the {\emph {statistical complexity}} $\Cmu$ of the \eM, 
given by~\cite{Crut89a,Shal01b}:
\begin{align}
    \Cmu = -\sum_{\sigma \in \mathbb{S}} \Pr ({\sigma}) \log_2 \Pr (\sigma)~.
\label{Eq:StatisticalComplexity}
\end{align}
$\Cmu$ is simply the Shannon entropy of the state probability distribution
and represents the average amount of memory (in bits) that the process
retains. As a general trend, the more CSs in an \eM, the larger $\Cmu$ and we
say that the process is more structurally complex.

More than one outgoing arc at a CS suggests that there is some uncertainty
about the next observed symbol. This notion of uncertainty can be quantified by
the {\emph {Shannon entropy rate}} $h_\mu$ and is directly calculable from the
\eM\ as~\cite{Shal01b}:
\begin{align}
   \hmu = -\sum_{\sigma \in \mathbb{S}} \Pr ({\sigma}) \sum_{s \in {\mathcal{A}}} {\sf{T}}^{(s)}_{\sigma \to \sigma^{\prime}}
                \log_2 {\sf{T}}^{(s)}_{\sigma \to \sigma^{\prime}}~.
\label{Eq:EntropyRate}
\end{align}
The ${\sf{T}}^{(s)}_{\sigma \to \sigma^{\prime}}$ are the probabilities for a
transition from CS $\sigma$ to CS $\sigma^{\prime}$ on symbol $s$~\footnote{When the ${\sf{T}}^{(s)}_{\sigma \to \sigma^{\prime}}$ 
  are written as $m \times m$ matrices, with $m$ being the number
  of CSs, these are the familiar {\emph {transition
  matrices}}~\cite{Riec14b} from the study of Markov models of stochastic
  processes. Also, note that due to \eM's unifilarity the symbol
  $s$ determines the unique destination state. And so, Eq.
  (\ref{Eq:EntropyRate}) does {\emph {not}} need to sum over $\sigma^\prime$.
  }.
The Shannon entropy rate gives the average uncertainty per measurement when 
all correlations are accounted for. It has units of [bits/measurement].  

While perhaps not obvious from casual examination, the \eMs\ in
Fig.~\ref{Fig:CrystalStructures} imply different Markov orders---the range of
interdependence. This is quantified by the {\emph {memory length}}
$\Range_{\ell}$~\cite{Varn13a}, an integer parameter that measures the maximum
range over which two symbol may carry nonredundant information about each
other. That is, there may exist correlations between symbols that are not
captured by the intervening symbols. It is possible that, even if the set of
states is finite, the memory length may
be infinite; these are the {\emph {strictly sofic}} processes~\cite{Weis73a}.

{\emph {Intrinsic computation}} is defined as how systems store, organize, and
transform historical and spatial information~\cite{Crut89a,Feld08a}. Different
processes may have quantitatively and qualitatively different kinds of
intrinsic computation, and understanding these differences gives insight into
how a system is structured~\cite{Crut94a}. In addition to the previous three
measures of intrinsic computation, there are others such as {\emph {excess
entropy}}~\cite{Crut83a}; {\emph {transient information}} and {\emph
{synchronization time}}~\cite{Feld04a}; {\emph{crypticity}}~\cite{Crut09a};
{\emph {bound information}} and {\emph {residual entropy}}; and {\emph {elusive
information}}~\cite{Jame11a}, each sensitive to different aspects of
information processing and storage. Usefully, it has recently been shown that
many of these information measures are directly calculable from the
\eM~\cite{Crut13a,Riec14a}. \\

{\bf \large{{Chaotic crystallography}}}

\begin{table*}[!htb]
\caption{({\it {left}}) A comparison of classical crystallography (\CLC) and chaotic crystallography (\CHC).   
               Notice the close parallels between the two descriptions. 
              ({\it {right}}) Measures of intrinsic computation for the \eMs\ in Fig.~\ref{Fig:CrystalStructures}
               and Fig.~\ref{Fig:MakeFigs}(b). The units of $\Cmu$ are bits, $\hmu$ are bits/ML, and 
               $\Range_{\ell}$ are MLs.  The abbreviations in the tables are: RT 3C = random twinned 3C; 
               RD 2H = random deformation 2H; NRD 2H = nonrandom deformation 2H; 
               NDT 3C = nonrandom (deformation and twinned) 3C.}
    \begin{minipage}{.5\linewidth}
      \centering        
       \begin{tabular}{|l|ll|}
       \hline
        \hspace{2.5cm}  & \CLC \hspace{1.9cm} & \CHC \hspace{2.0cm} \\
          \hline
           Material Structure          &  Crystal & Chaotic Crystal \\
                    &    &   \\
          Fundamental Unit         &  Basis / Unit Cell  &  Modular  Layers \\
            & &\\
         Organizational Schema     & Spatial Symmetry & Intrinsic Computation \\
         & &\\
        Mathematical Formalism & Group Theory & Semi-Group Theory \\
        & & \\
        Symmetries & Exact & Approximate \\
        & & \\
        Range of Applicability &  Crystalline  & Crystalline or Disordered \\
         \hline
      \end{tabular}
     \label{tab:word_probs_4B}        
  \end{minipage}%
  \begin{minipage}{.5\linewidth}
      \centering     
      \begin{tabular}{|lcccc|}
      \hline
      Example  & Material Structure & \,\,\,\, $\Cmu$ \,\,\,\, & \,\,\,\, $\hmu$ \,\,\,\, & \,\,\,\, $\Range_{\ell}$ \,\,\,\,   \\ 
     \hline
     1(a) &                 3C$^+$  & 0.00 & 0.00    &  0       \\
     1(b)  &              3C$^-$   & 0.00  & 0.00     &     0         \\
     1(c)  &             2H  &      1.00  &0.00 &  1            \\
     1(d) &            4H &         2.00 & 0.00 &   2                 \\
     1(e)  &            6H &         2.58 & 0.00   &   3            \\
     \hline
     1(f),    $q = 0.50$  & Random  &     0.00 &1.00  & 0 \\
     1(g), $\beta = 0.10$ & RT 3C & 1.00 & 0.47 & 1 \\
     1(h),    $\alpha = 0.10$ & RD 2H &1.00 & 0.47 & $\infty$ \\
     1(i),   $\eta = 0.10$ & NRD 2H &1.44 & 0.43 & $\infty$ \\
     2(b), SK137 & NDT 3C & 2.7 & 0.65 & 3 \\
     \hline
   \end{tabular}
   \label{Table:ComputationalMeasures}
  \end{minipage} 
  \label{Table:CHCandComputationalMeasures}
\end{table*}

{\emph {Chaotic crystallography}}
(\CHC)~\cite{Varn02a,Varn04a,Varn07a,Varn13a,Varn13b,Riec14b,Riec14c} is the
{\emph {application of information- and computation-theoretic methods to
discover and characterize structure in materials.}} The choice of the name is
intended to be evocative: we retain the term `crystallography' to emphasize
continuity with past goals of understanding material structure; and we
introduce the term `chaotic' to associate this new approach with notions of
disorder, complexity, and information processing. 

The idea of appealing to information theory to describe material structure is
not new, indeed Mackay has been a vocal and long-time proponent for such an
approach~\cite{Mack69a,Mack75a,Mack86a,Mack02a,Cart12a}. Until recently, though, a
comprehensive program to realize this vision was lacking. While \CHC\ does
realize this vision, it does not replace \CLC, but rather augments it,
providing a parallel, alternative view of structural organization in materials.
In many cases, especially for disordered materials, \CHC\ gives a more
consistent and comprehensive picture of material structure. We now show how
these information- and computation-theoretic tools can be incorporated in a new
view of material structure.

{\bf{Quasi-one-dimensional materials}}:
We specialize to the case where the periodic distribution of atoms is
preserved in two dimensions (2D), but not necessarily in the third,
as in the case of some {\emph {polytypes}} such as ZnS and SiC (they are isostructural)~\cite{Seba94a}.
A {\emph {modular layer}} (ML)~\cite{Ferr08a,Varn01a} is a sheet or plane
of atoms organized in a regular 2D array. For closed-packed structures (CPSs),
this is a hexagonal net. For ZnS in particular, at each lattice point in the net
there is a Zn-S pair, separated by one-quarter of 
a body diagonal (as measured along the conventional unit cell) in the direction 
perpendicular to the plane of the net, called the {\emph {stacking direction}}.
Since spatial symmetries are absolutely
respected within the MLs themselves, we can write the 2D version of equation~(\ref{Eq:CrystalStructure}) as:
\begin{align}
   {\mathrm {\bf{modular \,\, layer = basis}}} \times {\mathrm {\bf{2D \,\, lattice}}}. 
\label{Eq:ModularLayer}   
\end{align}


For CPSs, each ML can assume only one of three possible positions, usually
denoted \emph{A}, \emph{B} or \emph{C}, and
adjacent MLs stack according to the familiar closed-packed rule~\cite{Kitt05a}
that adjacent MLs may not have the same orientation. It is useful to take
advantage of this {\emph {stacking constraint}} and introduce the so-called H{\"{a}}gg notation, such that cyclic
transitions ($A \to B \to C \to A$) between MLs are labeled `1' and anticyclical
transitions ($A \to C \to B \to A$) are labeled `0'~\cite{Orti13a}. We define the
{\emph {stacking sequence}}~\cite{Varn02a} as the sequence of MLs encountered as one scans
the material along the stacking direction. The {\emph {stacking process}} is defined as the
effective stochastic process induced by sweeping the stacking sequence~\cite{Varn02a},
and we represent this in the H\"{a}gg notation over the binary symbols $\mathcal{A} = \{0,1\}$. 

Formally, for quasi-one-dimensional materials, \CHC\ divides the task of describing material structure into
two parts: (i) specify the structure of the fundamental unit, \ie, the (crystalline) 
2D MLs; and (ii) specify the mathematical construct that organizes the spatial distribution of 
the fundamental unit; \ie, the kind and amount of intrinsic computation as
captured by the \eM. 
The resulting material structure is referred to as a {\emph {chaotic crystal}}.
\CHC's analogous
relationship to \CLC's equation~(\ref{Eq:CrystalStructure}) is:
\begin{equation}
 \left (\begin{aligned}
      \mbox{} & {\mathrm {\bf{chaotic}}} \\
       \mbox{} & {\mathrm {\bf{crystal}}}
       \end{aligned}
 \right ) =
  \left (\begin{aligned}
      \mbox{} & {\mathrm {\bf{modular}}} \\
       \mbox{} & {\mathrm {\bf{layers}}}
       \end{aligned}
  \right ) 
\times
  \left (\varepsilonup\textbf{-machine} \right)
 ~.
\end{equation}
Notice the tight parallels between \CLC\ and \CHC: the material structure
(crystal versus chaotic crystal) is formed by taking a fundamental unit (basis
or MLs) and distributing it through space according to some mathematical
instruction (lattice or \eM). This close association between \CLC\ and \CHC\ is
summarized in Table~\ref{Table:CHCandComputationalMeasures} ({\emph {left}}).

{\bf{Methods for detecting intrinsic computation}}: 
Determining a material's intrinsic computation, by calculating or estimating
the \eM, is a primary goal of \CHC, and several methods have been explored in
the literature. Additionally, the causal architecture of the \eM\ provides invaluable information about the
stacking process, and this is explored in the examples shortly.  (i) One
method to obtain the \eM\ is to postulate causal architectures based on theoretical grounds. Estevez
\etal~\cite{Este08a} considered combined random growth and deformation faulting
in closed-packed crystals, and were able to generate a model that included
both, called the {\emph {random growth and deformation faults}} (RGDF)
process~\cite{Este08a,Riec14b}. Although this model is not unifilar, and thus
not an \eM, many of the techniques developed here can be adapted to analyze
it~\cite{Riec14a,Riec14c}. (ii) Another, statistical method is to simulate chaotic
crystals, and use one of the reconstruction methods available in computational
mechanics, such as the {\emph {subtree merging method}}~\cite{Crut89a}, {\emph
{causal state splitting reconstruction}}~\cite{Shal02a} or {\emph {Bayesian
structural inference}}~\cite{Stre14a}, to find the appropriate
model~\cite{Varn04a}. (iii) Lastly, the approach that has received the most
attention is {\emph {$\varepsilon$-machine spectral reconstruction theory}
(\eMSR)~\cite{Varn02a,Varn07a,Varn13a,Varn13b}.  The importance of this
technique is that it uses experimentally obtained X-ray diffraction patterns to
reconstruct the stacking process \eM.\\

{\bf \large{{Examples}}}

{\bf{Periodic stacking sequences}}: \CLC\ is well suited to describe periodic
stacking sequences. Being periodic, spatial symmetries are strictly obeyed, and
crystal structures are often specified using the Ramsdell notation $nX$, where
$n$ refers to the period of the repeated stacking sequence and $X$ to the
crystal system~\cite{Orti13a}. Commonly encountered crystal systems for CPSs
include the cubic (C), hexagonal (H) and rhombohedral (R). Examples are 3C$^+$
($\dots ABCABC \dots$), 2H ($\dots ABABAB \dots$) and 6H ($\dots ABCACB \dots$)
or in the H{\"{a}}gg notation these are ($\dots 111111 \dots$), ($\dots 101010
\dots$) and ($\dots 111000 \dots$), respectively. 

\CHC\ describes these familiar crystalline stacking structures in
the form of an \eM. For example, the 3C$^+$ stacking structure is compactly
given in Fig.~\ref{Fig:CrystalStructures}(a): an \eM\ with but a single CS and
a single transition. The 2H stacking structure,
Fig.~\ref{Fig:CrystalStructures}(c), is slightly more involved: there are a
pair of CSs connected by a pair of transitions. More involved still is the 6H
stacking structure, Fig.~\ref{Fig:CrystalStructures}(e), requiring six CSs and
six transitions. Indeed, for each of the first five \eMs\ in
Figs.~\ref{Fig:CrystalStructures}(a - e), each CS allows only one outgoing
transition, and the \eM\ describes periodicity. It should be apparent that any
such periodic repetition of CSs generates some crystal structure and that
crystal structures can only come from this kind of causal architecture. Closed,
finite, nonself-intersecting, symbol-specific paths on an \eM\ such as these
are referred to as {\emph {causal state cycles}}, and they are often specified
by putting in square brackets [$\cdot$] the sequence of causal states visited.

The measures of intrinsic computation defined in \CHC\ quantify crystal
structure and organization. Intuitively, we expect that the 6H is more complex
than say 3C$^+$ and indeed, by direct application of
equation~(\ref{Eq:StatisticalComplexity}), we find the statistical complexities
to be $\Cmu^{(\mathrm{6H})} = 2.58$ bits and $\Cmu^{(\mathrm{3C}^+)} = 0$ bits.
Thus, as we might expect on purely physical grounds, the 6H stacking structure
requires more computational memory than
3C$^+$.  Additionally, we observe that for each of these three examples, direct
calculation of the Shannon entropy rate using equation~(\ref{Eq:EntropyRate})
finds that $\hmu^{(\mathrm{3C}^+)} =  \hmu^{(\mathrm{2H})}  =
\hmu^{(\mathrm{6H})} = 0$ bits/ML, as we would expect for perfect crystal
structures. Lastly, we might imagine that somehow the 6H stacking structure
requires coordination between MLs at a greater length than that of either the
3C$^+$ or 2H stacking structures.  This notion is captured by the memory
length, and we find that for these three structures,
$\Range_{\ell}^{(\mathrm{3C}^+)} = 0$ ML,  $\Range_{\ell}^{(\mathrm{2H})}  = 1$
ML, and $\Range_{\ell}^{(\mathrm{6H})} = 3$ ML, confirming our intuition.

{\bf{Nonperiodic stacking sequences}}: When one moves beyond periodic stacking
sequences, strict symmetries are no longer maintained, but instead are
approximate. Mathematics based in the language of semi-groups---specifically
\eMs---is therefore more suitable than that of groups, which describe strict
symmetries.

We begin with a pedagogical example. Suppose that the stacking of MLs is random, in the
sense that other than respecting the CPS stacking constraints, there is no correlation
between MLs. If we allow for a bias in the stacking process---\ie, $\Pr(0) \neq
\Pr(1)$---then
the process is described as being {\emph {independent and identically distributed}} (IID)~\cite{Cove06a}. 
This process has been studied, for example by Guinier~\cite{Guin63a}, as a simple model of disorder. 
The \eM\ for the IID process is shown in Fig.~\ref{Fig:CrystalStructures}(f). One notes
a striking similarity with two of the periodic processes, namely the 3C$^+$ and
3C$^-$ in Figs.~\ref{Fig:CrystalStructures}(a) and (b). The one free parameter in the IID process is $q \in [0,1]$,
and adjusting it lets one scan from $q=1$, giving
a 3C$^+$ stacking structure, to $q=1/2$,  giving an entirely disordered structure, to lastly
$q=0$, giving the crystal structure 3C$^-$. From a \CHC\ point of view then, {\emph {the crystal
structures 3C$^+$ and 3C$^-$ are nothing more than special cases of a general IID model}} and
{\emph {this same IID model can also generate completely disordered stacking structures.}} This is perhaps
the clearest illustration of how perfectly crystalline and disordered materials may be computationally
similar. However, although they share nearly identical causal architectures, measures of intrinsic computation
do distinguish them. While we find $\Cmu^{(\mathrm{random)}} = \Cmu^{(\mathrm{3C}^+)} = 0$,
echoing their identical computational requirements; we also find $\hmu^{(\mathrm{random)}}(q=1/2) = 1.0$
bit/ML $\neq \hmu^{(\mathrm{3C}^+)} = 0$. This also illustrates the ease with which
\CHC\ can seamlessly encompasses both crystalline and disordered structures.

{\bf{Random versus nonrandom stacking faults}}: Many technologically useful
materials, such as SiC and GaP, are subject to stacking faults (SFs). And,
considerable effort is expended to characterize and understand them, often with
the intention of avoiding them during manufacturing. Let's see how the \eMs\ in
the last three panels of Fig.~\ref{Fig:CrystalStructures}---(g), (h), and
(i)---characterize various SFs in CPSs. 

The \eM\ in Fig.~\ref{Fig:CrystalStructures}(g)
represents the {\emph {random growth fault}} (RGF) process. For $\beta$ large, the RGF usually
oscillates between the two CSs, ${\mathcal{S}_0}$ and ${\mathcal{S}_1}$, giving 2H crystal structure.
With some small probability, an additional $1$ or $0$ is inserted into the stacking sequence and, physically,
this corresponds to a {\emph {growth fault}} of the 2H structure~\cite{Este08a}. At the other extreme
when $\beta$ is small, the RGF usually transits the state self-loops on each of the CSs and, physically,
repetition of each of these loops gives one of the 3C stacking structures. 
(Compare with Figs.~\ref{Fig:CrystalStructures}(a) and (b)). We recognize this as the 3C stacking structure
with randomly distributed twin faults. And, as we saw before, \CHC\ connects these two 
chaotic crystal structures (2H with random growth faults and twinned 3C) into a common causal architecture,
the only difference being in the transition probabilities.
Transformations between 2H and 3C are observed in ZnS~\cite{Seba94a} and, while a more
complex causal architecture is needed to describe the transformation, we see that in principle
\CHC\ provides very simple models to transform from one crystal structure to another.

The \eM\ in Fig.~\ref{Fig:CrystalStructures}(h) represents the {\emph {random deformation faulting}} (RDF) process
as it models random deformation faults in the 2H crystal structure~\cite{Este08a}. 
The introduction of deformation faults in 2H crystals is often modeled by 
{\emph {Glauber dynamics}}~\cite{Kabr88a,Varn04a} that corresponds to changing
$1$ to $0$ or $0$ to $1$. The \eM\ for the RGF does this randomly, with some small probability $\alpha$.

The \eM\ in Fig.~\ref{Fig:CrystalStructures}(i) is similar to the previous one, since for small $\eta$ it
too represents deformation faulting in the 2H structure, but now the SFs are
distributed {\emph {nonrandomly}} through the stacking sequence. 
We call this the {\emph {Nonrandom deformation faulting}} (NRDF) process. It is a simplified 
version of a previous model obtained from simulation experiments of the 2H $\to$ 3C transformation
in ZnS~\cite{Varn04a}. The critical difference between the RDF and the NRDF processes is the 
addition of two CSs `on the wings' of the 2H CSs---$\mathcal{U}$ and $\mathcal{V}$. These
extra CSs have the effect of preventing sequences that have an {\emph{even}} number of
$1$s or $0$s. Physically this implies that the occurrence of one deformation fault {\emph {suppresses}} the
occurrence of an adjacent deformation fault, and this is observed in
experiment~\cite{Seba87a}. Also, 
as the fault parameter $\eta$ grows, the chaotic crystal becomes increasingly dominated 
by odd-length sequence domains of $1$s and $0$s. Thus, this \eM\ reflects that the chaotic
crystal transforms into a nonrandomly twinned 3C crystal. 

\begin{figure*}
\begin{center} 
    \includegraphics[width=0.9\linewidth]{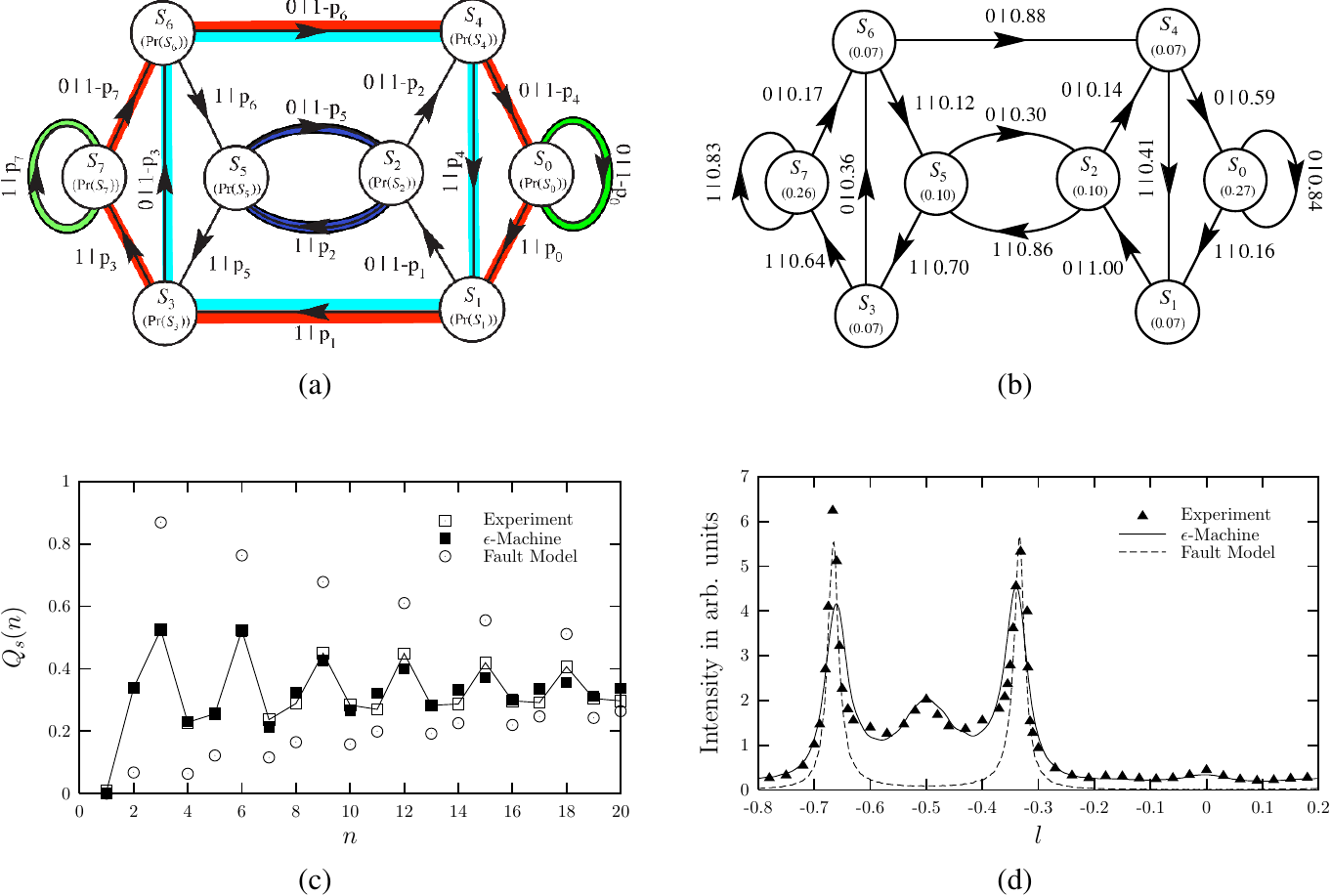} 
\end{center}    
    \caption{
        (a) The most general $\Range = 3$ \eM, with several of the more common 
        causal state cycles shown in color: 
        in green, [$\mathcal{S}_7$] and [$\mathcal{S}_0$] give the 3C$^+$ and 3C$^-$ crystal structures, respectively; 
        in blue, [$\mathcal{S}_5\mathcal{S}_2$] gives 2H; 
        in cyan, [$\mathcal{S}_3\mathcal{S}_6\mathcal{S}_4\mathcal{S}_1$] gives 4H; and
        in red, [$\mathcal{S}_7\mathcal{S}_6\mathcal{S}_4\mathcal{S}_0\mathcal{S}_1\mathcal{S}_3$] gives 6H.               
        (Adapted from Varn \etal~\cite{Varn13a}, used with permission.) 
        (b) The \eM\ that results from \eMSR\ when the experimental diffraction pattern ($\blacktriangle$) 
        in panel (d) is analyzed. (From Varn \etal~\cite{Varn07a}, used with permission.)
       (c) A comparison of the pairwise correlation function between MLs as obtained from 
           experiment ($\square$), the reconstructed \eM\ ($\blacksquare$) and an alternative
           description of the disorder, the fault model ($\odot$).  The $Q_{\mathrm {s}} (n)$ are the
    probabilities that two MLs at separation $n$ have the same absolute orientation (either $A,B$ or $C$.)
    The correlation functions are only defined for discrete values of $n$, and the line connecting
    adjacent points serves as an aid for the eye.
          (From Varn \etal~\cite{Varn07a}, used with permission.)
    (d) Comparison of the diffraction pattern calculated from the reconstructed \eM\ (---) and the fault model (- - -)
          to the experimental diffraction pattern ($\blacktriangle$). 
         (From Varn \etal~\cite{Varn07a}, used with permission.) 
    }
    \label{Fig:MakeFigs} 
\end{figure*}

Here then, we see two important points: (i) the causal architecture of the \eMs\ for chaotic crystals 
can sensitively reflect the structural organization of the stacking process; and (ii) the \eM\
seamlessly connects apparently different kinds of stacking processes into a single
causal architecture, facilitating the study of solid-state phase transitions. 
A major task in \CHC\ is the interpretation of the \eM\ in terms of physical
mechanisms that result in observed stacking processes. 

{\bf{\EM\ Spectral Reconstruction Theory}}: A significant source of information about crystals is X-ray diffraction, and
\CHC\ has a method of discovering intrinsic computation from this source, much as
\CLC\ uses X-ray diffraction studies to determine crystal structure. \eMSR~\cite{Varn02a,Varn07a,Varn13a,Varn13b}
employs Fourier analysis over a unit interval in frequency space to extract information about the pairwise correlations between
the MLs and then solves a set of equations for sequence probabilities. The algorithm initially 
considers low-order Markov processes and compares the diffraction pattern calculated from
the model with the experimental one. If the agreement is unsatisfactory, the order of the Markov model
is increased, and the comparison is repeated. This incremental process has been accomplished up to
third-order Markov models. The most general order-$3$ Markov model is shown in Fig.~\ref{Fig:MakeFigs}(a).

The triangles ($\blacktriangle$) in Fig.~\ref{Fig:MakeFigs}(d) show the diffraction pattern (corrected for
experimental effects) along the 10.$\ell$ row
of an as-grown disordered specimen of ZnS~\cite{Varn07a}. The degraded Bragg
reflections at $\ell \approx  -0.67$ and $\ell \approx -0.33$ are highly suggestive of
twinned 3C structure, but there is also considerable diffuse scattering, especially in the region
near $\ell \approx -0.5$. This is where one would expect to observe a Bragg reflection if 2H structure
were present, suggesting that in the disorder there may be some stacking sequences reminiscent
of 2H character. \eMSR\ was performed on this diffraction pattern over the interval $\ell \in [-0.80, 0.20]$,
and the resulting reconstructed \eM\ is shown in Fig.~\ref{Fig:MakeFigs}(b). The diffraction pattern
calculated from the reconstructed \eM\ is shown in a solid line (---) in Fig.~\ref{Fig:MakeFigs}(d) as
well as the diffraction pattern calculated from a competing model of disorder, the fault model~\footnote{
               We do not discuss the fault model in detail here, but we note that this model
               is based on \CLC, where one assumes a perfect crystal `corrupted' by some fault structure~\cite{Varn02a}. 
               While often useful for weakly faulted specimens, it is not tenable when the disorder is 
               large, such that the crystallographic symmetries are appreciably broken.}, in a dashed 
line (- - -). Clearly, the \eM\ is successful in capturing the broadband scattering near $\ell \approx -0.5$
and it also reproduces the Bragg-like reflections near $\ell \approx  -0.67$ and $\ell \approx -0.33$,
though the peak intensities are somewhat less than that observed in experiment. From other processes
reconstructed from diffuse diffraction patterns, it is known \eMSR\ can sometimes have
difficulty faithfully reproducing the line profiles~\cite{Varn07a}. 

The correlation function for the probability that two MLs at separation 
$n$ have the same absolute orientation (either $A$, $B$, or $C$)
extracted from the experimental diffraction pattern ($\square$) are shown in 
Fig.~\ref{Fig:MakeFigs}(c), along with those from the reconstructed \eM\ ($\blacksquare$)
and an alternative description of the disorder, the fault model ($\odot$). 
For small $n$, the agreement between the correlation function calculated from the
\eM\  and experiment is rather good, but becomes less so at larger $n$. One explanation
for this is that there are correlations between MLs that a third-order Markov model
has difficulty reproducing. Indeed, simulation studies on solid-state phase transitions
in ZnS~\cite{Varn04a} suggest that no finite-order Markov model is capable of
exactly capturing all the structure. 

Examining the reconstructed \eM\ in Fig.~\ref{Fig:MakeFigs}(b) we observe the high
state probabilities for the CSs $\mathcal{S}_0$ and $\mathcal{S}_7$ as well as their
large self-loop transition probabilities, confirming that this is a twinned 3C crystal, albeit with considerable disorder.
Notably, the next most visited CSs are $\mathcal{S}_2$ and $\mathcal{S}_5$, and they do have a
relatively small but nonetheless nonnegligible inter-state transition probability between them.
This causal state cycle would give 2H crystal structure, if it were more strongly represented.
So, there does seem to be some 2H character in the stacking process, although it is weak.
The remaining states represent transitions between these two structures; \ie, they are faulting
structures. For highly disordered specimens, such as this one, it is often difficult 
to unambiguously assign a particular fault or crystal structure to specific architectural
features~\cite{Varn02a,Varn07a} and a more nuanced investigation, coupled with simulation
studies is required. It is clear that for many real crystals, however, that the disorder can be profound and
not as simply represented as the processes of Fig.~\ref{Fig:CrystalStructures} might imply.

This is an example of the kind of analysis that is possible with \CHC. Close coupling between experimental
investigations, simulation studies, and theoretical reconstruction procedures is promising as
a highly effective tool for discovering, characterizing, and explaining disordered stacking structures.\\

{\bf \large{{Future directions}}}

While \CHC\ is still in its infancy, it has potential to significantly impact the way disordered
structures are understood, discovered, and described. Since the modeling procedure is based 
in the mathematics of (probabilistic) semi-groups, it can naturally accommodate inexact or approximate symmetries
such as those found in disordered materials, where \CLC\ loses applicability. 

Future directions include expanding on recent developments in understanding
spectral properties of \eMs~\cite{Crut13a,Riec14a,Riec14b,Riec14c}, where they can
be a powerful quantitative tool. In particular, calculating material properties, such as
thermal and electronic transport through disordered media via their \eM\ representation, offers a
way to systematically search the space of disordered processes for interesting and useful
phenomena. Additionally, measures of intrinsic computation, so closely linked as they are to
structure, are likely to strongly correlate with material properties. 

Another research direction is applying \CHC\ to materials in higher dimensions; \ie, treating
2D materials. Although the formalism as reviewed here concentrated on quasi-one-dimensional
materials, the basic notions transfer to higher dimensions, and this is an area of current research.

Lastly, we return to one of the initial motivations of crystallography, as encapsulated in the
question we began with---{\emph {Where are the atoms?}} \CLC\ gives an unambiguous answer in the form
the material's crystal structure. In its use of probabilities, it seems perhaps that \CHC\ has failed to reach this goal. The
answer offered by \CHC, however, is at once both new and informative in a different way.
\CHC\ finds and examines 
the process that describes the material, and this may not only be a more convenient,
but a more insightful answer. From the process, computational and physical parameters are calculable; 
and the space of possible configurations is given a kind of order, permitting systematic investigation.
This is because \CHC\ does not necessary tell where each and every atom is (although it does in
the case of periodic processes), but rather it defines an ensemble of configurations, as specified by
the \eM, that statistically represents the material. And often, this is enough.\\

{\bf{{Acknowledgments}}}

The authors thank Julyan Cartwright, Chris Ellison, 
Alan Mackay, John Mahoney, Tara Michels-Clark, Paul Riechers and 
Richard Welberry for comments on the manuscript 
and the Santa Fe Institute for its hospitality during visits. JPC is an SFI External Faculty member. This
material is based upon work supported by, or in part by, the U.S. Army Research
Laboratory and the U. S. Army Research Office under contract W911NF-13-1-0390.

\bibliography{ChemOpinion}

\end{document}